\documentclass[notitlepage, reprint,onecolumn,superscriptaddress]{revtex4-1}
\usepackage{amsmath}
\usepackage{graphicx}
\usepackage{bm}
\usepackage{natbib}
\usepackage{mathtools}
\usepackage{subfig}

\def\mnras{Mon.Not.Roy.As.Soc.}

\def\ltsima{$\; \buildrel < \over \sim \;$}
\def\simlt{\lower.5ex\hbox{\ltsima}}
\def\gtsima{$\; \buildrel > \over \sim \;$}
\def\simgt{\lower.5ex\hbox{\gtsima}}

\def\m@th{\mathsurround=0pt }
\def\eqalign#1{\null\,\vcenter{\openup1\jot \m@th
 \ialign{\strut\hfil$\displaystyle{##}$&$\displaystyle{{}##}$\hfil
 \crcr#1\crcr}}\,}

\def\[{\begin{equation}}
\def\]{\end{equation}}

\newcommand{\ud}{\mathrm{d}}

\newcommand{\mean}{\mu}  
\newcommand{\pop}{N}

\newcommand{\rank}{r}
\newcommand{\nU}{N_t} 
\newcommand{\median}{M}  
\newcommand{\medianI}{\median_I}  
\newcommand{\medianG}{\median_G}  

\newcommand{\rhalf}{r_h}  
\newcommand{\ntotal}{N}
\newcommand{\births}{B}

 \newcommand{\nmax}{{N_{\mathrm{max}}}}

\newcommand{\dist}{\bm{\mathcal{G}}}  
\newcommand{\draw}{N}  

\newcommand{\minpop}{\pop_m}  

\newcommand{\ensemble}{\dist} 
 
\newcommand{\ele}{\epsilon}  

\begin{document}
\title{Apocalypse Now?  \\ Reviving the Doomsday Argument}
\author{Fergus Simpson}
\email{fergus2@icc.ub.edu}
\affiliation{ICC, University of Barcelona (UB-IEEC), Marti i Franques 1, 08028, Barcelona, Spain.}


\begin{abstract}
Whether the fate of our species can be forecast from its past has been the topic of considerable controversy. One refutation of the so-called Doomsday Argument is based on the premise that we are more likely to exist in a universe containing a greater number of observers. Here we present a Bayesian reformulation of the Doomsday Argument which is immune to this effect. By marginalising over the spatial configuration of observers, we find that any preference for a larger total number of observers has no impact on the inferred local number. Our results remain unchanged when we adopt either the Self-Indexing Assumption (SIA) or the Self-Sampling Assumption (SSA).  Furthermore the median value of our posterior distribution is found to be in agreement with the frequentist forecast.  Humanity's prognosis for the coming century is well approximated by a global catastrophic risk of $0.2\%$ per year.
\end{abstract}

\maketitle

\section{Introduction}
Few of us believe we have a high chance of celebrating our 100th birthday, yet alone our 200th. But what's the source of our pessimism? Why do you suspect that your death will occur within the next few decades, if this is an event that you've never previously experienced? Acceptance of our own mortality follows from a simple observation: each one of us appear to be an ordinary human individual. As such, we are able to learn from the experiences of others, to gain information about our individual selves.  Most of our personal attributes are already well known to us, so if we were to measure how common a particular eye colour or hair colour is among the population,  our state of knowledge is not improved. But if we happen to be ignorant of a particular trait, such as our lifespan, then the population distribution does serve as a valuable source of information. For example, I don't know my blood type, but I consider my chances of having the type AB-  to be less than $1\%$, on the basis that less than $1\%$ of the population has AB- blood.

These inferences regarding your life expectancy, and the outcome of your blood test, may appear fairly innocuous.  We are making informed predictions about future events, based solely on the past experience of other individuals. In this respect, the Doomsday Argument is no different. The premise of the Doomsday Argument is that we have no particular reason to believe we were born especially early or late in the context of all human lineage \citep{carter1983anthropic, gott1993implications}. It would therefore be surprising to find ourselves among the first $1\%$ of all humans who ever exist, in exactly the same way that it would be surprising to find any particular one of our physiological features - such as the level of haemoglobin in your blood - to be in the first percentile. Couple this belief with an approximate understanding of the number of humans who have already been born, and this suggests we can construct an estimate for the total number of births we expect to occur in the future.

While the finiteness of our personal future is uncontroversial \footnote{Though many may cling to the hope that miraculous advances in biotechnology might render them immortal.},  forecasting the fate of our species has proved to be much more divisive. Various attempts have been made to evade the conclusions of the Doomsday Argument \citep{dieks1992doomsday, 1994Kopf, 2000Olum}. The most widely accepted refutation stems from the assertion that we have a greater probability to exist in a more highly populated universe. This has been claimed to tip the balance in favour of a lengthy future for the human race. Yet this would lead us - and everyone else who adopts this algorithm - to believe that we live in an unusually early period within the lineage of human history. This is a paradoxical conclusion - how could the majority of people all belong to a small minority?

There are several scientific motivations for resolving this conundrum. The selection effects associated with our existence have a broad range of potential applications, from the inference of cosmological parameters \cite{1989RvMP...61....1W, 2006PhRvD..73b3505T}, to understanding the features of our galaxy \cite{2014RaulGRBs} and our planet \cite{2016SimpsonOceans}. They may also prove invaluable in directing the search for extra-terrestrial life \cite{2015SimpsonAliens, simpson2016development} \footnote{See also the excellent animation by MinutePhysics: \url{https://www.youtube.com/watch?v=KRGca_Ya6OM}}. Yet the most important consequence is not scientific, but geopolitical. If we were to reach a consensus that the threat of global catastrophe is significant, then preventative measures are more likely to be enacted \cite{rees2004our, bostrom2011global}. 

In this work we present a Bayesian reassessment of the Doomsday Argument, with the aim of resolving the paradoxical conclusions mentioned above. In particular, we shall account for the substantial uncertainty in the spatial distribution of observers in the universe.   In \S \ref{sec:typical} we begin by considering the reference classes within which we should consider ourselves to be a representative sample.  In \S \ref{sec:groups}  we consider the implications of drawing a single sample from an array of groups of  unknown size. The generic result for this puzzle is then applied to the case of observers in \S \ref{sec:doomsday}, where we present our main results. A comparison with previous work, including a revised assessment of the Fermi Paradox, can be found in \S \ref{sec:comparison}, while our concluding remarks are in \S \ref{sec:conclusions}.

\section{Typical Observers} \label{sec:typical}

The notion of typicality is one where, in the absence of additional information, one ought to reason as if you were selected at random from a broader population. This is something we all invoke  intuitively   when we wish to estimate our life expectancy or our blood type. In this section we shall aim to put the concept of typicality on firmer footing, rather than relying on intuition, by taking stock of the available evidence.  \citet{2007Hartle}   question whether typicality is justified, making the following claim:
\begin{quotation} 
``At present, there are no observational data supporting an assumption that we are typical in some class of observers"
\end{quotation}
Yet  such evidence can be acquired in just a few minutes.  
\subsection{Establishing Typicality}  \label{sec:typicality}
First we  shall \emph{empirically} establish our typicality among one particular class - humans - before employing a thought experiment in order to extend typicality to observers on other worlds.

\begin{itemize}
\item{\textbf{STEP 1} - Collate a list of your personal data and compare each with the global population. These can include country of birth; blood type; hair colour; and so on. Do you find yourself falling into those categories which have higher populations? }

This analysis should lead you to conclude that you are an ordinary human, whose properties are reflective of the global distribution. Without accepting this step, it is difficult to justify the following statement:  "I am more likely to have a common blood type than a rare one". 

\item{\textbf{STEP 2} - If you suffered from a severe bout of amnesia and couldn't recall your country of birth, how would you assess the relative probability among different countries? }

In accordance with step 1, it ought would be weighted by the number of births in that country, for which we can use population size as a proxy. Selecting any other weighting scheme would imply a preference towards one nationality over another. It would also lead to a prior which is sensitive to the secession and unification of countries.

\item{\textbf{STEP 3} - If mankind had colonised other planets, how would you assess the relative probability of having been born on these different planets?}

These colonies are just new countries, albeit at a slightly more distant location.  In accordance with step 2, each planet must be weighted by population size, in the same way that countries are.

\item{\textbf{STEP 4} - What if those other colonies had not travelled from the Earth, but had evolved independently?}

If we reach the same final configuration by a different means, it is difficult to justify why our inference ought to change. We should therefore consider ourselves typical among some broader class of observers, if they exist. This is a direct consequence of  the empirical evidence gathered in Step 1. 
\end{itemize}

\subsection{The Failure of Neutrality} 
 
\begin{figure}
\includegraphics[trim=4cm 2cm 3cm 2cm, clip, width=75mm, angle = 90]{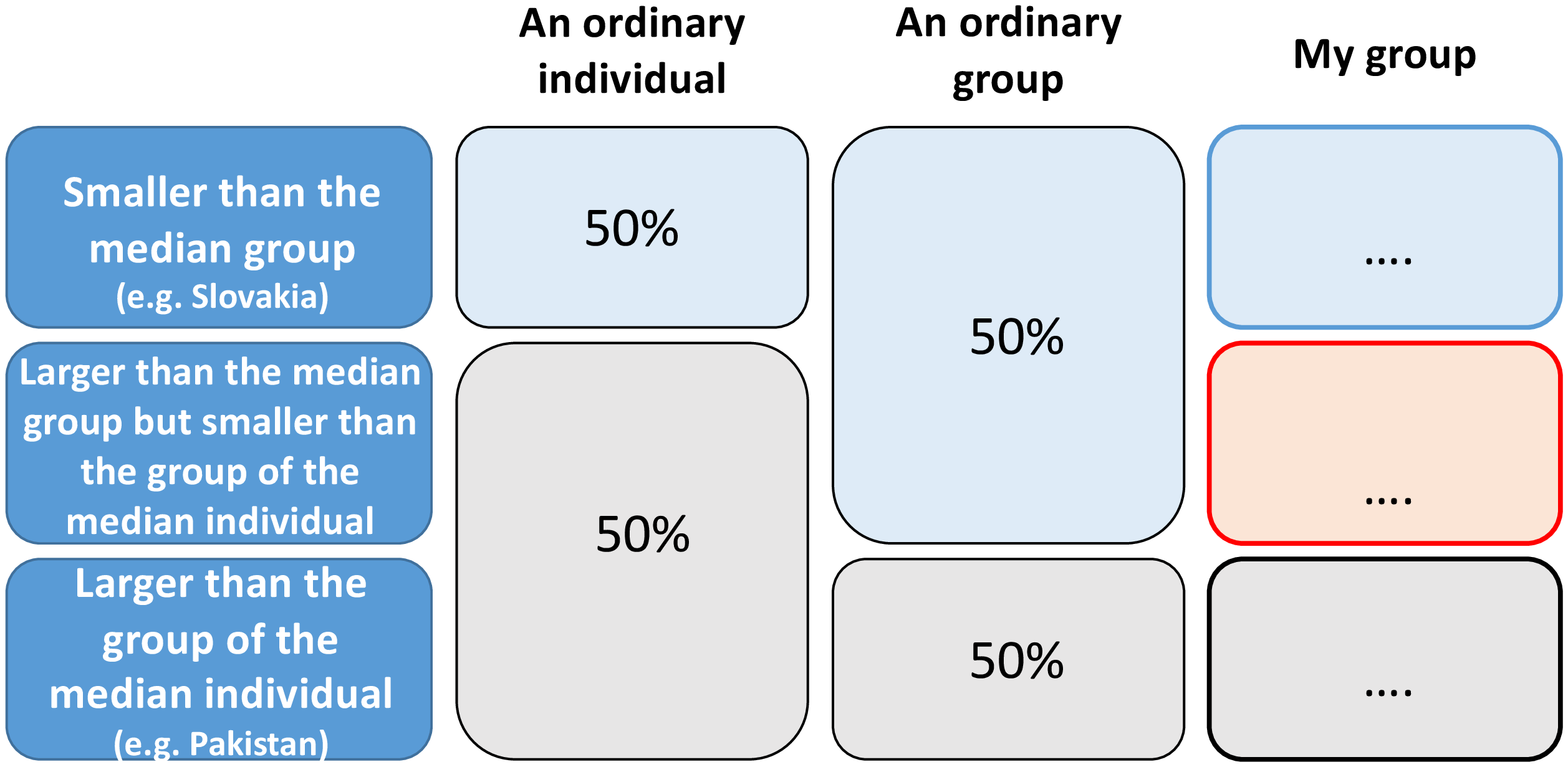}  
\caption{ A demonstration of why one cannot maintain a neutral belief for both yourself and the group to which you belong. The group in question could be your blood type, nationality, or civilisation. In order to satisfy both conditions one must leave the central panel at $0\%$. This is clearly inappropriate, unless all groups are the same size.  
 \label{fig:medians}}
\end{figure}

One might still be tempted to reject typicality, and instead insist that the most conservative belief is one of blissful ignorance.  Wouldn't it be safest to assume that we actually know nothing, and all possibilities remain equally likely? Figure \ref{fig:medians} demonstrates how such a belief system is not logically self-consistent. If one insists on adopting a state of pure agnosticism, that would suggest that you believe that our group of observers is equally likely to be a higher population or lower population than other groups of observers. In other words, the probability that our group lies on either side of the median group size $\medianG$ is $0.5$. So we must enter $``50\%"$ in the upper panel in the right hand column. But equally, to remain truly agnostic, this logic must apply for our belief relative to other individuals. That is to say, we are equally likely to lie in a larger or smaller group than most other individuals, who experience a different median group size $\medianI$. So we must then also enter $``50\%"$ in the lower panel in the right hand column. But then, in order to ensure a normalised  probability distribution, we are obliged to enter $``0\%"$ in the central panel.  Therefore, by attempting to declare total ignorance, we have arrived at a nonsensical conclusion. There must be some finite probability of lying between the median individual and the median group.  To suggest otherwise is equivalent to asserting that all the groups are of identical size  \cite{2015SimpsonAliens}.   
   
The distinction between the median group  and the group of the median individual,  is often overlooked, because they are of a similar magnitude when the different groups are all of a similar size. However, for ensembles of populations with larger variances, these quantities quickly diverge, such that $\medianI \gg \medianG$. A case in point is the contrasting populations of Slovakia (5.4 million, currently ranked 117 of 233 states by population size) and Pakistan (192 million, currently ranked sixth, but it is the country of the median individual, in that an equal number of people reside in larger and smaller countries). Clearly the gap between these two benchmark populations is not negligible - over $48\%$ of the world's population live in countries which fall between them.

\subsection{Reference Classes}
If we can consider ourselves to be typical among some class of observers, how should we define the limitations of the group?  Adopting different boundaries creates different reference classes, and this can lead us toward diverging conclusions. This has previously  been the  cause of some concern \cite{bostrom1999doomsday}.  To see why such ambiguity is completely benign, we proceed with the following worked example. Let's rank every human in accordance with the latitude at which they were born. We begin by assigning a rank of 1 to whoever was born closest to the South Pole, a rank of 2 to the second closest, and so on.  This provides each of us with a spatial ranking, rather than the temporal one used in the Doomsday Argument. In order to draw a more direct comparison with the temporal ranking (where we  only have knowledge of one `direction' - the past) let us also imagine that your only knowledge of human and physical geography is south of your place or birth, so you have no idea what lies north of that position, or how many people live there. I was born at a latitude of approximately 56$^\circ$N, and I could choose to designate myself to one of many reference classes. This includes worldwide observers, Europeans, British, Scottish, or those born in that particular hospital.  The existence of multiple classes is in no way problematic. Each one permits a credible exercise in statistical sampling \cite{2008Garriga}. Within each and every class, my rank (i.e. how many observers within that class were born further south than I was) serves as a reasonable estimate of how many were born north of my position. Hence  this allows me to estimate the total population within each class. The reader is encouraged to verify this calculation with their own data (which can be repeated using your current location in lieu of your place of birth).

If I were to deliberately select a landmark that is close to my current location - such as the Mediterranean Sea - there are only a small number of people between it and myself. I should \emph{not} then use this number to conclude that there are few people living north of my location. Doing so would be a classic example of the posterior selection fallacy.  Yet analogous arguments have been made with the Doomsday Argument. There have been relatively few individuals born since the advent of nuclear weapons, but this fact should not be used to forecast an imminent apocalypse. There will always be nearby `landmarks' in human history, and since we can only see those in the past, we can only identify those reference classes that we have recently fallen into. We can't identify the future ones we narrowly miss out on, such as the advent of strong AI. Our asymmetric perspective of these temporal reference classes has led to the misconception of a reference class problem. The strategy required to avoid falling into this trap is straightforward - do not select landmarks in an \textit{a posteriori} fashion, such as those in our recent history.  As is made clear from the latitude analogy,  the existence of multiple reference classes presents no barrier to drawing inferences on the total population within each class, provided they are constructed in an unbiased manner.


\begin{table*}
\begin{tabular}{|p{2 cm}|p{8cm}|p{5cm}|}
 \hline
 \bf{Symbol}   	& \bf{Variable}   &  \bf{ Definition} \\ 
 \hline
$\ele$   &  Element  &  \\ 
$\rank $  & Rank within the group &  \\  
$\pop$           &   \text{Population of the group}&  \\ 
$\dist $  &   Set of parameters defining the distribution p(N) &  \\
\hline
$\mean$     &        \text{Mean  population}     &  $   \int_0^\infty \pop p(\pop) \ud \pop $  \\ 
$\nU $  & Total number of elements in all groups &  $\sum_i N_i$ \\
$\medianI $  & Population of the group of the median element & $ \mu^{-1} \int_0^{\medianI} N p(N)  \ud \pop = 0.5 $   \\  
$\medianG $  & Population of the median group  &  $ \int_0^{\medianG} p(N)  \ud \pop = 0.5$   \\   
 \hline
\end{tabular}
\caption{Guide to the notation}
\label{tab:1}
\end{table*}

\section{Sampling from an unknown ensemble} \label{sec:groups}

Scientists across a broad range of disciplines are obliged to conduct experiments in a `blind' manner, meaning the true outcome remains hidden until the data analysis has been finalised.   This critically important piece of protocol ensures that the experimental results are shielded as much as possible from the unwanted influence of cognitive biases. For much the same reason, we strongly advocate that theoretical calculations with emotionally charged consequences (such as the fate of our species) ought to be performed with dehumanised examples where possible. Only once satisfied with our answer, should we proceed to connect the analogy to the problem at hand.

Consider a finite array of urns, where the $i$th urn holds a total number of balls $N_i$. The local ensemble of urns defines a probability distribution $p(\pop)$, where $\pop$ is the number of balls held by a single urn. The distribution $p(\pop)$ shall be parameterised using the set of variables $\dist$.  Within each urn, every ball is individually labelled based on the order they were placed in the urn, from 1 to $N_i$.  All balls are poured into a single super-urn, from which a single ball is then drawn. This ball is found to be labelled with the number $\rank$. What does this single piece of information tell us about (a) the urn from which it originated, and (b) the urn population as a whole?

\subsection{Inference on the chosen urn}
 
\begin{figure*} 
\includegraphics[width=55mm]{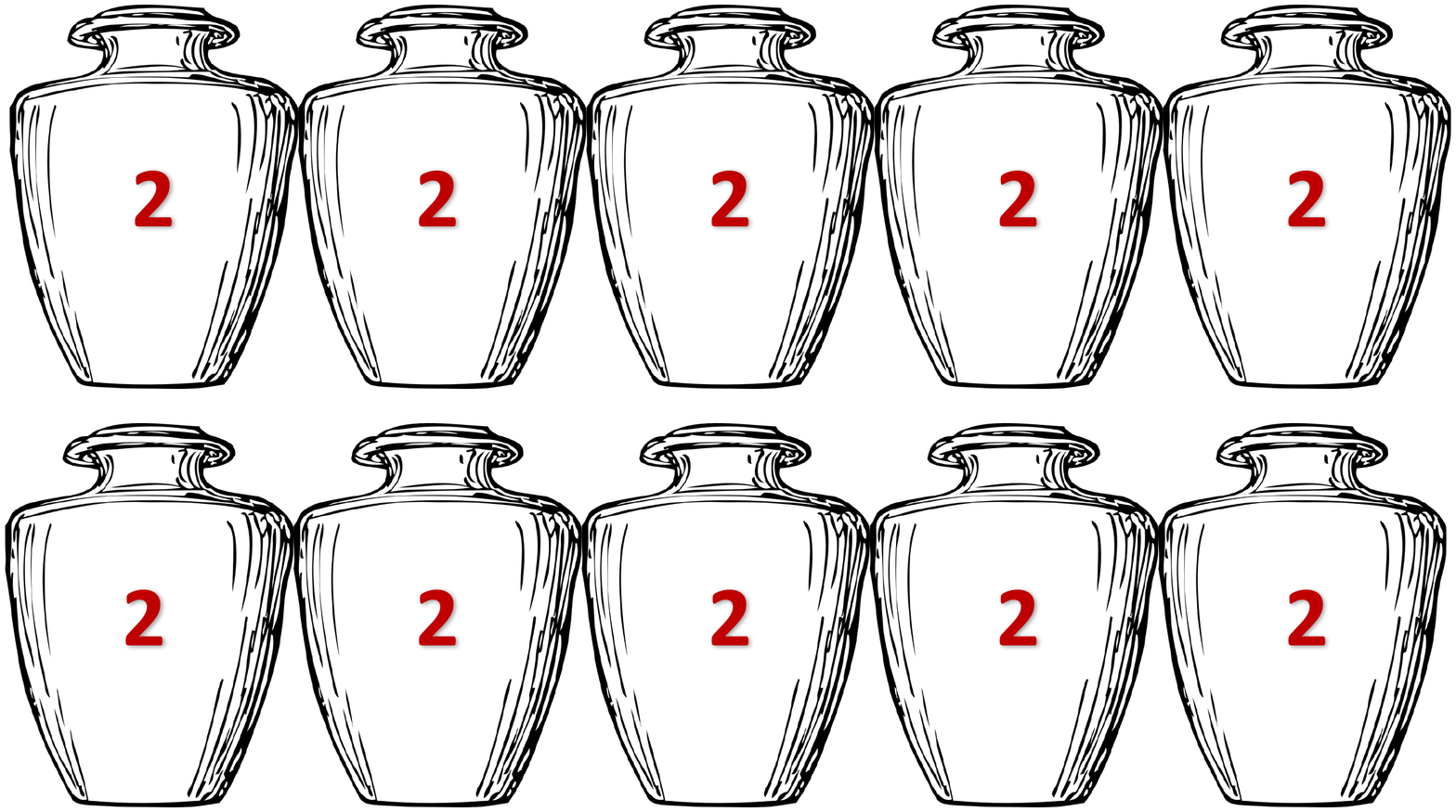}
\includegraphics[width=55mm]{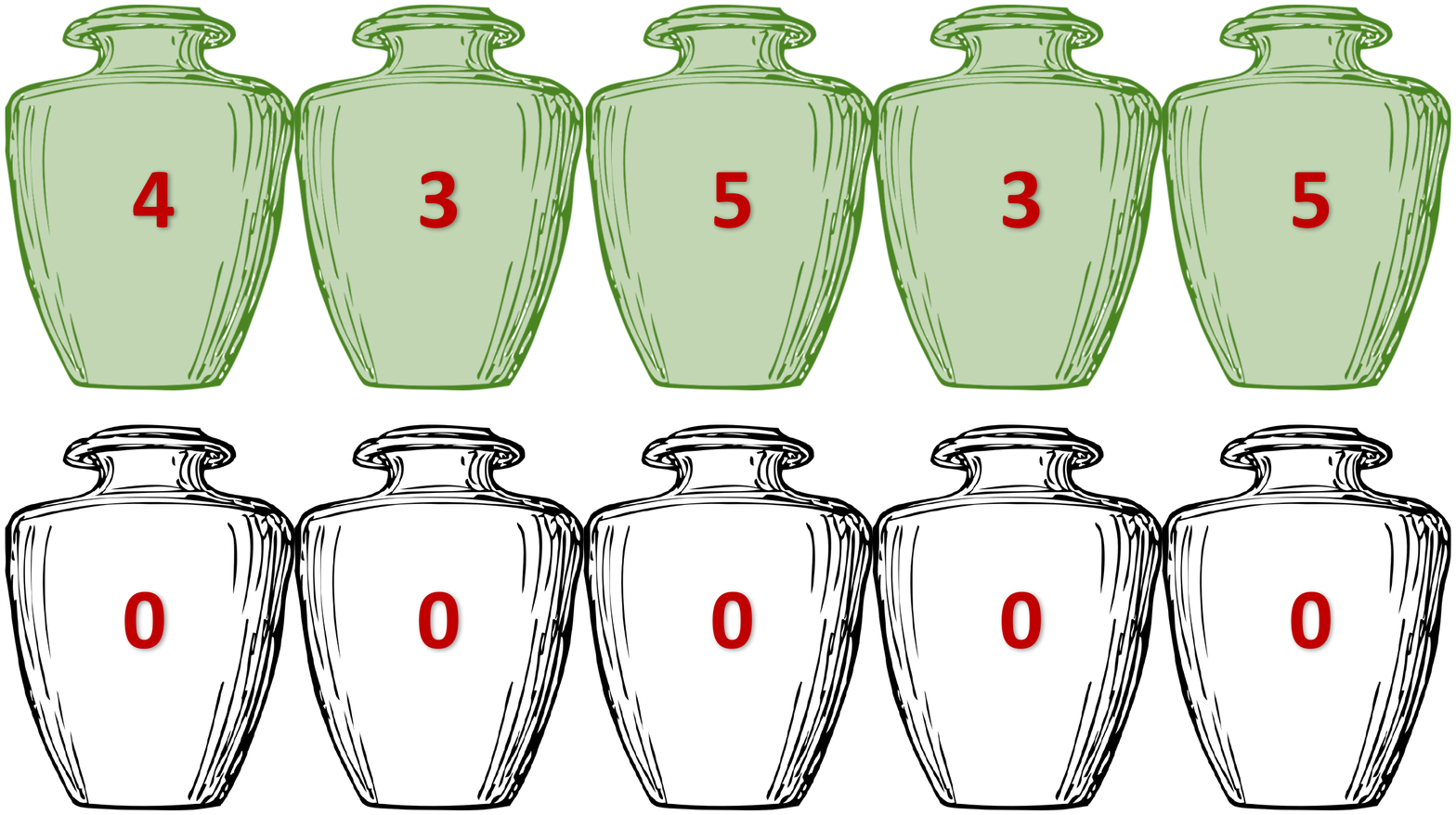}
\includegraphics[width=55mm]{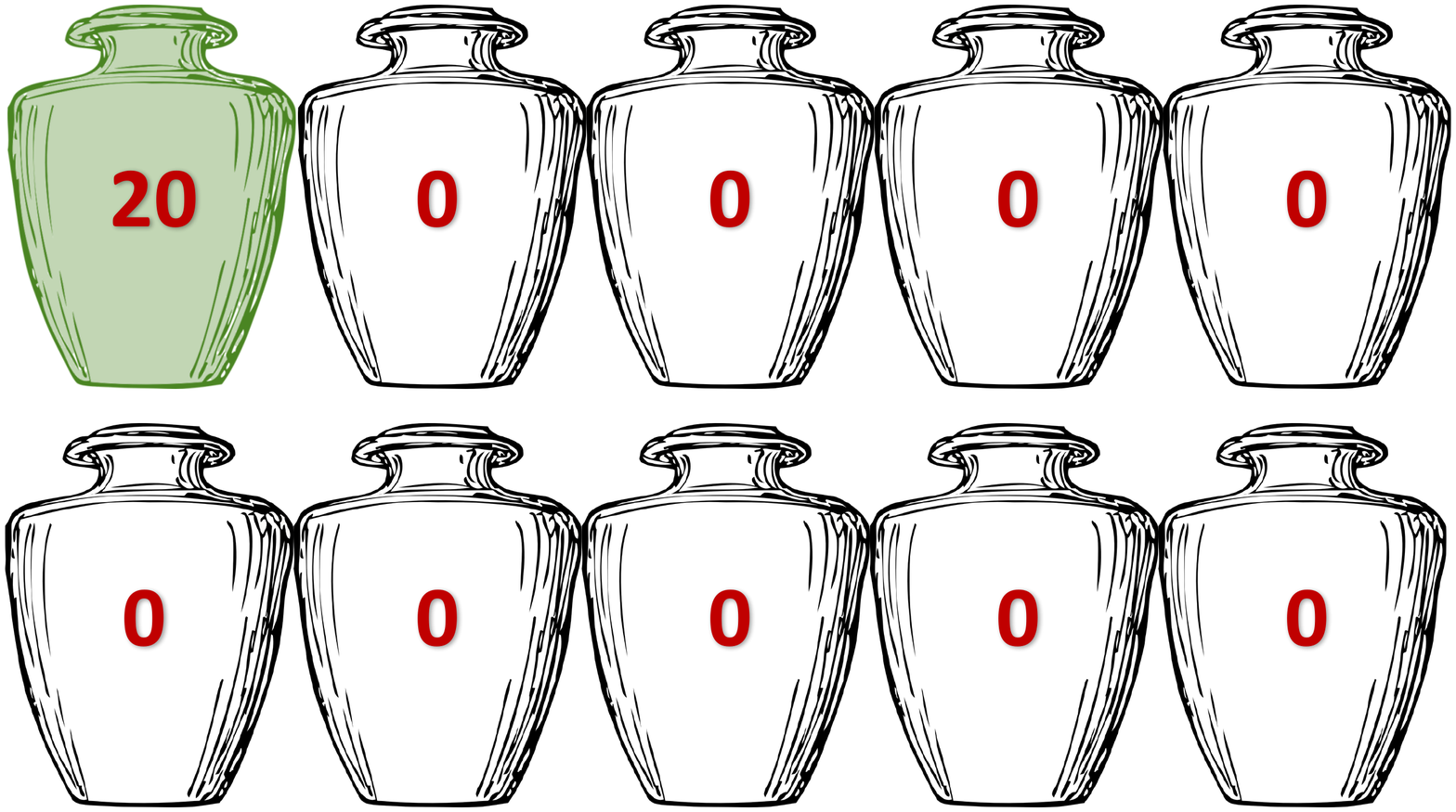}
\caption{Twenty numbered balls are distributed across ten urns, with the numbering reset within each urn.  There are known to be three candidate distributions, illustrated above. A single ball is drawn from the global distribution and found to be labelled ``3".  As a result of this observation, we can (a) eliminate the candidate on the left, and (b) favour the second distribution over the third, with odds of $5:1$.} \label{fig:urns}
\end{figure*}

Figure \ref{fig:urns} illustrates a simple example of the more general problem under consideration. In this particular case our prior $\pi(\ensemble)$ involves only three candidate distributions. In each possible scenario there are 20 balls spread across ten urns. If we draw a single ball labelled $`3`$, then the candidate on the left can be eliminated. Meanwhile the likelihood of the central distribution is five times greater than the likelihood of the right hand distribution, since it contains five times as many balls labelled $`3'$.  

With the aim of developing a more general solution, let us begin by treating the total number of balls, $\nU \equiv \sum N_i$, as a fixed quantity, while the number of urns is unknown. For simplicity we shall also work in the limit of large numbers, $\nU \gg \rank \gg 1$, such that the discrete distributions may be well approximated as continuous. Given some prior belief for the local ensemble $\pi(\ensemble)$, we can express the probability distribution for the number of balls in the original urn $\pop$, selected by a single ball $\ele$, as follows

\[
\eqalign{ \label{eq:pN}
p(\pop |\rank, \ele) &=  \int  p(\pop |\rank, \ele, \ensemble) \pi(\ensemble) \ud^n \ensemble  \, , \cr
&\propto \int  p(\rank | \pop, \ele, \ensemble) p(\pop | \ele, \ensemble) \pi(\ensemble)\ud^n \ensemble  \, .
}
\]
The ball-selected distribution for $N$, $ p(\pop | \ele, \ensemble)$, is related to the true urn distribution $ p(\pop | \ensemble)$ as follows
\[
 p(\pop | \ele, \ensemble) = \frac{\pop p(\pop | \ensemble)}{\int \pop' p(\pop' | \ensemble) \ud \pop'}
 \]
since the chance of selecting a particular urn is proportional to the number of balls it holds. For a given value of $N$, the probability of selecting a ball labelled $r$ is
 \[
  p(\rank  | \pop) =
  \begin{dcases} 
       \frac{1}{\pop}   ,& \pop \geq r \, ,\\
    0,              & \pop < \rank \, ,
\end{dcases} 
\]
Substituting these relations into (\ref{eq:pN}), we arrive at the general expression for the posterior on $\pop$, 
\[ \label{eq:posterior}
 p(\pop |\rank, \ele) \propto
\begin{dcases} 
       \int \frac{p(\pop | \ensemble)}{\mean(\ensemble)}  \pi(\ensemble) \ud^n \ensemble  ,& \pop \geq \rank \, ,\\
    0,              & \pop < \rank \, .
\end{dcases} 
\]
where $\mean(\ensemble)$ is the mean value of $N$ across the ensemble, as defined in Table \ref{tab:1}. The impact of observing a single rank $\rank$ is therefore twofold:

\begin{enumerate}
\item{ It eliminates the possibility $\pop < \rank$  }
\item{Our prior belief is updated to favour those ensembles with lower mean population sizes.}
\end{enumerate}

Why should we favour  ensembles with a smaller mean group size? The probability of selecting a ball with a value $\rank$ is proportional to the total proportion of $\rank$-labelled balls. And if some urns held a very large population,  far exceeding  $\rank$, this ensures that only a very small proportion of all the balls can be labelled $\rank$.
 
Finally we can generalise our result to the case where the total number of balls $\nU$ is unknown.   Note that $\nU$ makes no explicit contribution to the posterior probability given by (\ref{eq:posterior}). So marginalising over $\nU$ is now trivial 

\[
\eqalign{ 
 p(\pop |\rank, \ele) &= \int  p(\pop | \rank, \ele, \nU) p(\nU)  \ud \nU  = p(\pop | \rank, \ele, \nU)  \int p(\nU) \ud \nU \, ,  \cr 
&=   p(\pop | \rank, \ele, \nU)      \, .  \cr  
}
\]
Therefore,  provided $\nU \gg \rank$, the inferences we draw are insensitive to the prior $p(\nU)$. 
 
\subsection{Inference on the urn distribution}

What impact does this observation of $r$ have on our belief of the local ensemble $\dist$?  We can immediately eliminate those distributions  for which $p(\pop \geq r) = 0$. In other words, we now know that at least one urn holds at least $r$ balls.  But this is not the only implication, as we can see from the relationship between the posterior $p(\ensemble | r)$ and our prior belief $\pi(\ensemble)$. 
 
 \[
\eqalign{ 
p(\dist |r) &\propto  \pi(\dist) p(\rank | \dist)  \, , \cr
& \propto   \frac{\pi(\dist)}{\mean(\dist)}  \int_r^{\infty} p(\pop | \dist) \ud \pop \, .
}
\]
The probability of the \emph{local} distribution function, defined by the set of parameters $\dist$, is therefore penalised if either (a)  many urns possess a population size smaller than $r$,  or (b) too many balls are occupying urns far exceeding the rank, as this ensures that there are very few balls labelled $\rank$. 

\begin{figure}
\includegraphics[width=100mm]{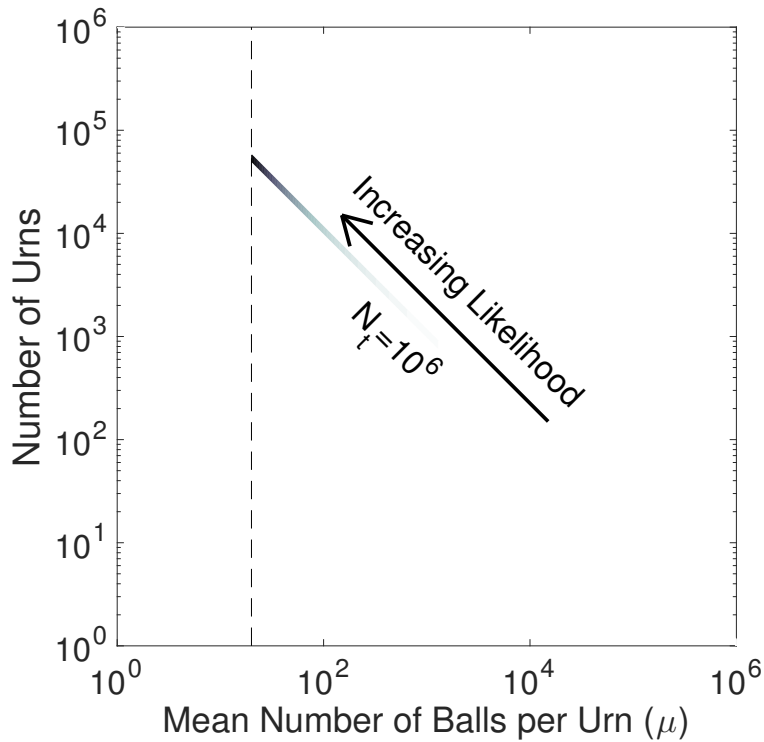}
\caption{ An illustration of the higher likelihood associated with smaller populations.  In this simplified example, one million balls are uniformly distributed across an unknown number of urns. A single ball is drawn from the full ensemble, with a value of $\rank = 20$. This observation leads us to strongly favour those configurations which involve a large number of urns which hold at least 20 balls.
\label{fig:dist}}
\end{figure}

Figure \ref{fig:dist} depicts a worked example.  For simplicity we enforce a uniform distribution, so $\dist$ only consists of a single parameter, the mean $\mean$. The total number of balls is set at $10^6$, and a single value of 20 is drawn. Those models with a larger number of urns and correspondingly  smaller $\mu$ have a much higher likelihood, scaling as $\mu^{-1}$  until the point where $\mu < 20$, at which point the likelihood vanishes.  Note that this preference for $\mu \simeq 20$ holds for any value of $\nU$, so as emphasised above, there is no sensitivity to our choice of prior $p(\nU)$.

\section{Implications for Humanity} 
\label{sec:doomsday}

We now proceed to connect the urn scenario with the spatial distribution of observers in the Universe.  Each ball corresponds to an observer,  while each urn corresponds to their planet (or more generally, astronomical body) of origin.  The observer can establish their typicality via the procedure outlined in Section \ref{sec:typicality}. Furthermore it should be noted that the ensemble can be interpreted at a local level, or equally it could encompass all possible observers, in accordance with the Self-Indication Assumption \cite{bostrom2003doomsday}. This is defined as

\begin{quote}
\it{An observer should reason as if they are randomly selected from the set of all possible observers.}
\end{quote}
\noindent where the random selection is weighted by the likelihood of the scenario. 

\subsection{Marginalising over the ensemble of ensembles}
 
In order to marginalise over a family of distributions $\dist$, we must decide on the functional form to describe the populations sizes of civilisations in our universe. One attractive option, based on its success in modelling population sizes, is to use the Pareto distribution. It reflects a power law distribution, coupled with a minimum group size $\minpop$. Previous applications of the Pareto distribution include the populations of cities \cite{rosen1980size}, microbial species \cite{hong2006predicting}, and Bose-Einstein condensates \cite{ijiri1975some}. The probability distribution associated with the Pareto distribution is as follows

\[ \label{eq:pareto}
p(\pop) = 
\begin{dcases} 
     \frac{\alpha \minpop^\alpha}{\pop^{\alpha+1}}   ,& \pop \geq \minpop \, ,\\
    0,              & \pop < \minpop \, ,
\end{dcases} 
\]
where the Pareto index $\alpha$ lies within the range $1 < \alpha < \infty$.  The mean of the distribution is given by
\[
\mean = \frac{\alpha \minpop}{\alpha - 1}  \, .
\] 

To evaluate the posterior distribution on $\pop$, given the observation of a single rank, we make use of the general result given by (\ref{eq:posterior}). The set of parameters $\dist$ is now comprised of $\alpha$ and $\minpop$, such that
\[
\eqalign{
 p(\pop |\rank, \ele) &= \int_0^\infty \! \! \! \int_1^\infty  \frac{1}{\mean}   p(\pop | \alpha, \minpop)   \pi(\alpha, \minpop) \, \ud \alpha \, \ud \minpop \,, \cr
               &=  \int_0^N \! \! \! \int_1^\infty  \left( \frac{\alpha - 1}{\alpha \minpop} \right)  \frac{\alpha \minpop^\alpha}{\pop^{\alpha + 1}} \pi(\alpha, \minpop) \, \ud \alpha \, \ud \minpop \,.
               }
 \]
Adopting Jeffreys' prior for the minimum group size $\pi(\minpop) \propto 1/\minpop$, and integrating over $\minpop$ we find
 
\[ 
p(\pop |\rank, \ele) \propto \frac{1}{\pop^2} \int_{1}^{\infty} p(\alpha)    \ud \alpha \, ,
\]
for   $\pop \geq \rank$. Remarkably, we can choose any prior on the Pareto index $\alpha$, and we will arrive at the same result, specifically

\[  \label{eq:humans}
p(N|r) = 
\begin{cases}
    \displaystyle{\frac{\rhalf}{N^2}} ,& N \geq \rhalf \\
    0,              &  N <  \rhalf
\end{cases}
\]
where $\rhalf \equiv r - \frac{1}{2}$, and the subtraction of one half follows from approximating the discrete distribution as a continuous one.  That this result is insensitive to the prior on $\alpha$ is indicative of its broad applicability. 

\subsection{The Doomsday Forecast}

 To utilise our result given by (\ref{eq:humans}), we require an estimate of our rank, $r$. In other words, how many humans have there been to date? Motivated by the work of \citet{haub1995many}, we adopt a fiducial value of $r_0 = 10^{11}$. Clearly there exists a significant uncertainty in the true value of $r$, and we account for this by allowing $r$ to vary by a factor of three from its fiducial value.  Specifically, we set $p(\log r) = \text{constant}$ across the range $ \frac{1}{3} r_0  < r < 3 r_0$. 
This uncertainty may seem substantial, yet it has a minimal impact on our results. The effect is a very slight worsening of short term prospects, and a very slight improvement in the far future. 

\begin{figure}
    \centering
        \subfloat[The  probability distribution for the number of future individuals. ]{{\includegraphics[width=8cm]{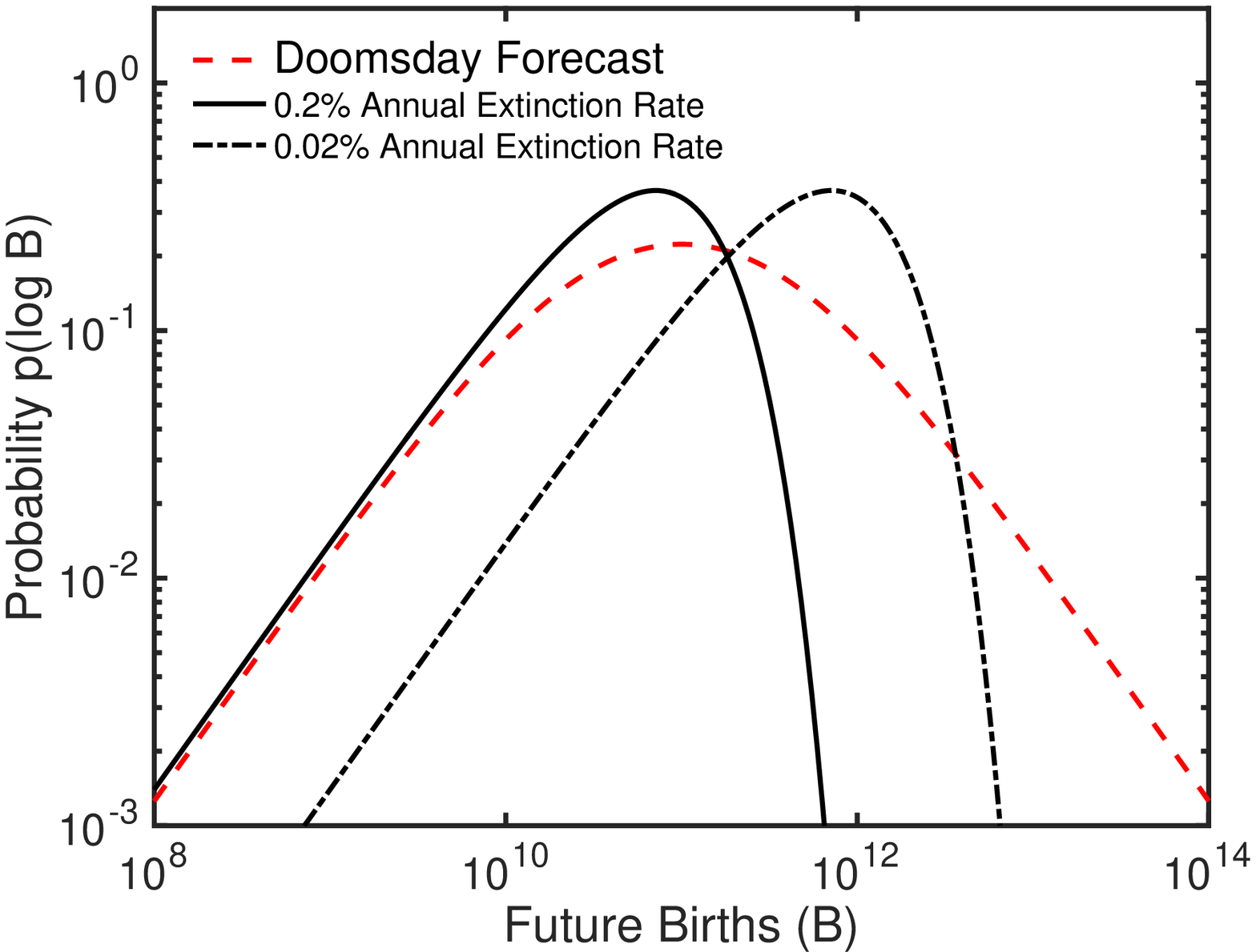} }}%
    \qquad
    \subfloat[The probability that human extinction will have occurred by a given date. ]{{\includegraphics[width=8cm]{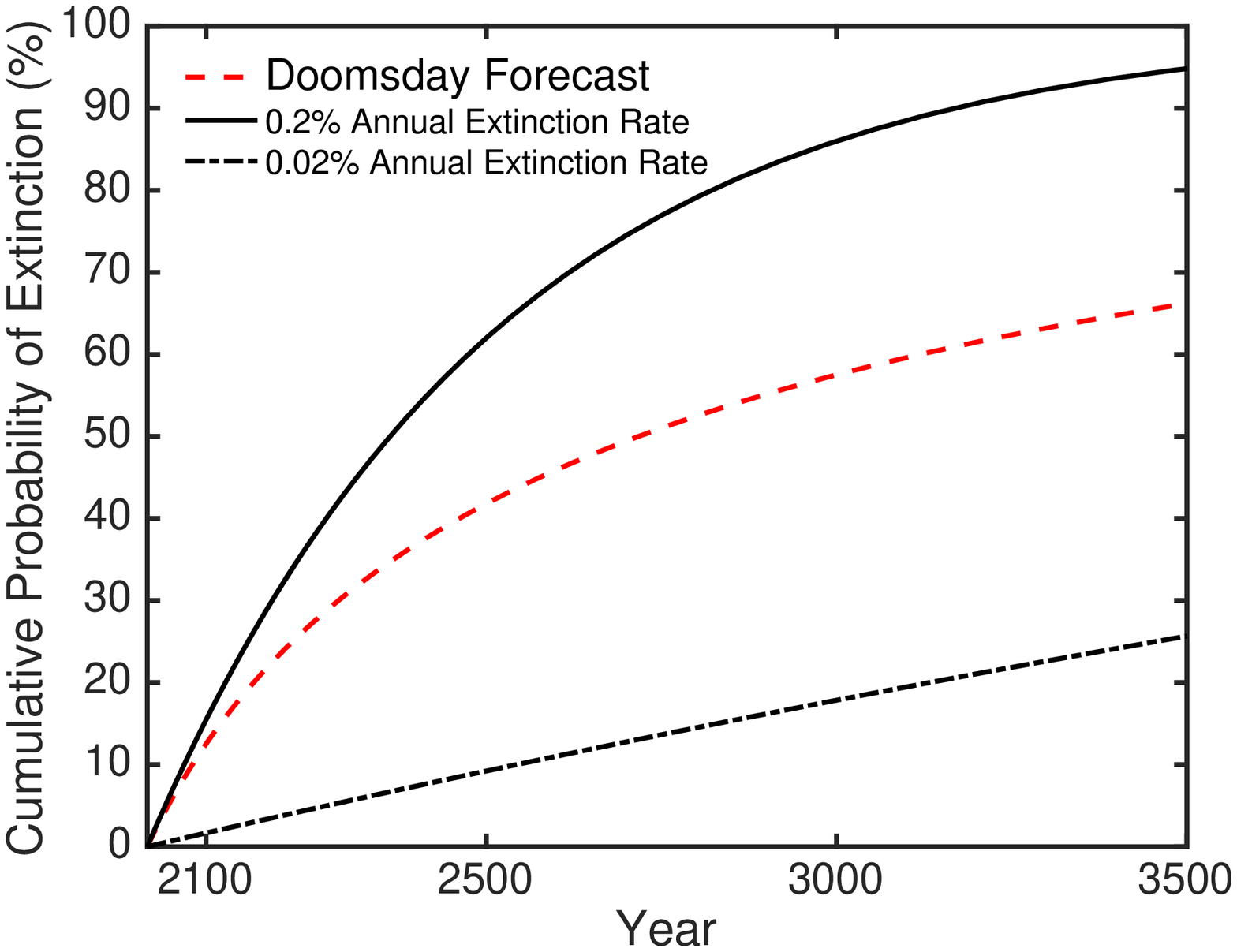}}}%
    \caption{In each panel, the dashed line represents the probability distribution derived from equation (\ref{eq:humans}), accounting for our uncertainty in $r$. For comparison, the solid and dot-dash lines represent models where the annual probability of extinction is fixed at $0.2\%$ and  $0.02\%$ respectively. When transposing the Doomsday forecast onto the right hand panel (and when transposing the temporal models onto the left hand panel) we assume that the number of births per year remains constant. The  number of observers per civilisation is modelled as a Pareto distribution, as defined in (\ref{eq:pareto}), and we have marginalised over the minimum population size $\minpop$.}
\label{fig:extinction}
\end{figure}

Our central results are presented in Figure \ref{fig:extinction}. In the left panel the dashed line represents our forecast for $\births$, the total future number of observers on Earth (we need not limit ourselves to those classified as \textit{Homo sapiens}).   This may be expressed analytically, for a given value of $r$, by recasting (\ref{eq:humans}) as follows

\[
p(\births |r) = \frac{r}{(\births+r)^2} \, .
\] 

In the right hand panel we reparameterise our forecast from the left hand panel into a function of time. In order to achieve this we require a model for the future global birth rate.  Over the past thirty years, this number has remained approximately constant, at close to 140 million per year \footnote{Source: \url{http://data.un.org}}. In the short term we can therefore be confident that this number will serve as a good estimate, barring a global catastrophe.  We find that, over the course of the 21st century, the Doomsday forecast is well approximated by an extinction risk of $0.2\%$ per year.   

How do these results compare with other assessments of our future prospects? For example, \citet{rees2004our} estimated a $50\%$ probability of human extinction by the year 2100. A study by \citet{2008globalrisk} arrived at a risk factor of $19\%$ over the same timeframe. The results of this work are only slightly more optimistic. Here we estimate a $13\%$ chance that humanity shall fail to see out the 21st century. This equates to odds of 7:1.

Pressing further into the future, the forecast becomes less reliable, since the birth rate could change significantly. Nonetheless, in this particular model, our civilisation's `life expectancy'  (the date we are equally likely to pass or fall short of) is approximately 700 years from the present. This appears consistent with an independent estimate of ${\sim}400$ years, which derives from the longevity of historical civilisations \cite{shermer2002hasn}. 

\subsection{The Total Number of Observers}

Why do we not appear to favour models which yield a larger total number of observers $\nU$, as some other calculations have claimed? A brief answer is that any given value of $\nU$ can not only be generated by the emergence of a single large group, but is equally likely to arise from the formation of a large number of small groups. And it is this latter configuration which is vastly more likely to yield the observed value for $\rank$. So a preference for a larger value of $\nU$ has no bearing on our inference of the group size $\pop$.  We shall revisit this issue in Section \ref{sec:comparison}.

Throughout this work we have considered  $\nU$ to be a finite quantity. Here we shall briefly consider the possibility that the number of observers is in fact infinite.  If the universe is spatially infinite (and there is currently no evidence to suggest otherwise), this will ensure an infinite number of civilisations. This offers the most viable route to forming an infinite number of observers, but this also has no clear impact on our calculation. What if the mean number of observers per civilisation, $\mu$, was infinite? This would require at least one civilisation to produce an infinite number of individuals. However this prospect seem much less likely, as it appears to violate the laws of thermodynamics. The act of cognition requires both a supply of energy (see Landauer's principle) and a corresponding rise in the total entropy. Each observer must therefore exploit some free energy. Yet within a given patch of de Sitter space, this is a quantity which is in finite supply.  The total number of observers generated by a single civilisation must therefore be finite. Thus, quite independent of our statistical considerations, our understanding of cosmology   also points towards a scenario in which a large number of observers are more readily generated by means of a diverging number of civilisations, rather than intrinsically large civilisations. 

\section{Comparison with Previous Work} \label{sec:comparison}

Our findings are in tension with the conclusions of several previous studies. In this section we shall explore possible sources of these discrepancies. 

\subsection{The SIA Correction}

Perhaps the most frequently cited rebuttal to the Doomsday Argument relates to the increased probability of existing in a universe with more observers, as summarised by the relation $p(N | I) \propto N p(N)$   \citep{dieks1992doomsday, 1994Kopf, 2000Olum}.  To begin we reproduce the following sequence which attempts to relate the posterior probability of our total group size $N$ given our observed rank $r$,

\[ \label{eq:SIAposterior}
\eqalign{
p(N|r) &\propto p(r | N) p(N | I) \, , \cr
 &\propto p(r|N) N p(N) \, , \cr
 &\propto  p(N) \, .
}
\]
The above relationship has been claimed to show that the prior is unaffected by the observation of $r$. However this procedure  neglects two critical points. First of all, for group sizes smaller than our rank, $N<r$, the posterior vanishes, $p(N|r) = 0$. An improved description of the posterior probability is therefore given by

\[  \label{eq:posteriorcase}
    p(N|r) \propto 
\begin{cases} 
    p(N) ,& N\geq r\\
    0,              & N < r
\end{cases} 
\]
This is  the single most important mechanism  underpinning the Doomsday Argument. In most cases, the size of a group in which we do exist will greatly exceed the typical size of any given group. This is simply because, as we have already established, we are much more likely to exist within the largest groups. So we should fully expect the truncation effect given by (\ref{eq:posteriorcase}) to eliminate the vast majority of \emph{any} proper broad prior $p(N)$.   

The second error in the logical sequence  of $(\ref{eq:SIAposterior})$ is the exploitation of an improper prior.   To ensure the $1/N$ prior is proper we must truncate the function at some finite value $\nmax$. In doing so this causes the observed group size $p(N | I)$ to stack up hard against this limit. For example if we take $\nmax$ to be $10^{100}$ we can immediately conclude with $99\%$ confidence that an observer's population falls within the narrow range between $10^{98}$ and $10^{100}$. This is clearly a pathological situation. Adopting a prior as informative as this prohibits the likelihood of the data from exerting a significant influence on the posterior.  As we have seen, a straightforward resolution is found if we permit variation in both the group size \emph{and the number of groups}.

\subsection{The Posterior Selection Fallacy}

It has been argued that the extreme improbability associated with our exact memories and DNA sequence suggests we should favour the existence of a large number of observers  \cite{2006math......8592N}.  As before, we advocate working with a dehumanised analogy in order to protect against cognitive bias.  In the United Kingdom, the National Lottery involves drawing a sequence of six numbered balls from a well mixed pot of 49 (recently increased to 59).  Imagine you only have a single piece of knowledge about the history of this lottery, which was that  the following set of balls has been drawn at some point in time

\[ \label{eq:lottery}
06 - 17 - 24 - 29 - 33 - 36 \, .
\]
We shall denote this data as $x$. Based solely on the above information, what was the total number of lottery draws $\draw$ which have been performed? One might be tempted to apply Bayes' Theorem as follows
\[
p(\draw  | x) \propto p(x| \draw ) p(\draw) \propto \draw p(\draw) \, .
\]
It would then appear much more likely that there were a large number of draws, limited only by the upper bound on the prior.  Yet it is clearly an inappropriate conclusion to reach based on the evidence presented in (\ref{eq:lottery}). No information on the number of draws is provided by this specific set of numbers, because these numbers were selected \emph{after the event}.  Conversely, if these numbers had been chosen in advance, then later we discover that they had come up, then this calculation does indeed provide evidence in favour of a large number of draws. 

The above pitfall is entirely analogous to the line of reasoning applied in the standard Doomsday rebuttal, which attempts to justify a large number of observers based on the tiny probability of our existence. It is therefore fallacious to exploit one's personal history and DNA to determine the total number of observers. One should not be overly surprised by the rarity of last week's lottery numbers, and in the same manner, one should not be surprised by the details of your own existence.

\subsection{Extreme Observers}

Some objections to the Doomsday Argument involve putting oneself in the shoes of an extreme observer, such as an individual living near the North Pole,  or someone who lived hundreds of years ago \cite{2004Smolin, 2006math......8592N}.  From this perspective, a Doomsday-like Argument is used to derive some spurious results, which are clearly in conflict with the truth. Should we be concerned by this?

When any statistical statement is made, we do so with a limited degree of confidence, such as $98\%$. We therefore fully expect to be wrong 2\% of the time. Those individuals who fall within the extremities of a given distribution, be it their latitude, altitude, height, or date of birth, simply represent those occasions where the prediction will fail, \emph{as it must}. If we construct 50 independent metrics by which we can rank the population, the chances are that you will be extreme in one of them, but the prediction will succeed on the other 49 occasions. 

Part of the confusion is that extreme observers make educated guesses that are several orders of magnitude away from the truth. Yet it is clear from the left panel of Figure \ref{fig:extinction} that the distribution has a highly extended tail. So it is not especially surprising to find that your group size is a few orders of magnitude greater than your rank.

\subsection{The Frequentist Calculation}

Next we compare our results to those derived from a frequentist perspective. Instead of considering different ensembles,  we fix the ensemble and vary our position within it. Let's use our rank $\rank$ as an estimator for the total group size, as follows. 

\[ \label{eq:freq_rank}
\hat{N} = 2 \rank - 1 \, .
\]
Taking the ensemble average over all observers within any given group yields
\[
\eqalign{
\langle \hat{N}  \rangle &= \sum_{r=1}^N  (2r - 1) p(r)   \, ,\cr
&=   N     \, .\cr
}
\]
We can therefore see that  $\hat{N}$ is an unbiased estimator of the group size, and this holds for any ensemble $p(N)$.  It is surprising that the arguments of \citet{gott1993implications} have sometimes been criticised as being based on intuition, when the mathematical justification is straightforward.  In any case,  how does this compare to the Bayesian result? The posterior distribution given by (\ref{eq:humans}) has a median value of
\[ \label{eq:median}
\median (\ntotal) = 2 r  - 1\, .
\]
Despite these two very different methodologies, both the Bayesian and frequentist calculations agree that $(2 \rank -1)$ represents our best guess for the size of our group.

\subsection{The Fermi Paradox}

 \begin{figure*} 
\includegraphics[width=160mm]{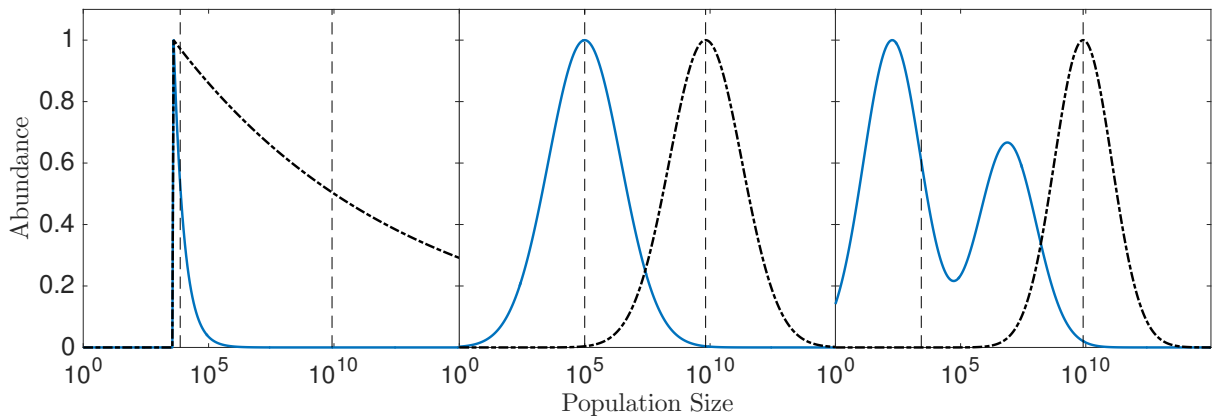}
\caption{Three toy models for the distribution of population sizes among civilisations. In each panel, the solid line denotes the true ensemble, while the dot-dashed line represents the distribution as sampled by individual observers. In each case, the gap between the two vertical dashed lines highlights the vast difference between the median group $M_G$ and the group of the median individual $M_I$. From left to right, we see examples of the Pareto, lognormal, and bimodal lognormal distributions. In each example the distribution parameters   are selected such that the median individual inhabits a population of 7 billion.} \label{fig:civs}
\end{figure*}

The Fermi Paradox questions why we have yet to hear from another civilisation, if intelligent life in the galaxy is plentiful. This has inspired the proposal of numerous solutions \cite{2009SerCirkovic}, including the rarity of extra-terrestrial life \cite{barrow1986anthropic}; the self-destructive nature of civilisations; our inability to recognise other civilisations; that they are listening but not transmitting; and that they are deliberately avoiding contact \cite{ball1973zoo}. While several of these points may hold some degree of truth, they miss the bigger picture. The entire debate is founded on the notion that our civilisation is a fair representation of other civilisations. Yet when introspectively studying one's own civilisation, an individual observer gains an immensely distorted perspective of what other civilisations may be like.  In the following we shall estimate the magnitude of this bias, before considering its implications. 

Let's begin by considering the distinction between the population of the median civilisation $\medianG$, and the population to which the median observer belongs $\medianI$. The extent to which $\medianI$ exceeds $\medianG$ is governed by the degree of diversity among population sizes. If the variance is large, then the gulf between the two will be vast,  $\medianI \gg \medianG$. And a large variance seems to be unavoidable, due to the temporal variations in our own population size, and also the large number of contributing variables. To estimate the magnitude of this effect, Figure \ref{fig:civs} depicts three mock realisations. In each panel the solid line represents the true distribution, while the dot-dashed line is the distribution as sampled by individual observers. From left to right, the three functional forms are: (1) a Pareto distribution with $\minpop = 4,000$ based on measurements of the minimum viable population \cite{traill2007minimum}; (2) a lognormal; and (3) a bimodal lognormal. In each case the model parameters are chosen such that the median observer belongs to a population of $\medianI = 7 \times 10^{9}$. The likelihood of each model is therefore high.  Despite these different functional forms, there is a familiar outcome in each case. Only a very small proportion of civilisations exceed $\medianI$: of the order $10^{-6}$, $10^{-4}$, and $10^{-3}$ respectively. Furthermore, in all three models the median population size is less than one million ($\medianG < 10^{6}$).  This may sound surprisingly small, yet this is consistent with the populations of other terrestrial species \footnote{All species with sufficient body mass to house brains of similar size to our own, have populations $N \simlt 10^6$.}.  
 
These toy models offer a glimpse into the observational bias we are susceptible to, when we extrapolate our population to those of other intelligent species. Once this bias is taken into consideration, it becomes clear that the overwhelming majority of civilisations are unlikely to have the resources required to develop an extensive space program. It is no coincidence that the only nations capable of crewed spaceflight have populations in excess of 100 million (namely Russia, China, and formerly the USA).  Furthermore, these nations have all benefited from global technological progress.  Such global progress may be dramatically slower in civilisations with greatly reduced populations, and much scarcer resources.

Putting aside the special status of our civilisation, there is a further compelling reason why the lack of contact is both unsurprising and uninformative. Several authors have insisted that we ought to have found tangible evidence of other civilisations, such as self-replicating  Von Neumann (VN) probes \cite{simpson1964view, hart1975explanation, tipler1980extraterrestrial, barrow1986anthropic}.   
Yet even if technological civilisations like ours are in abundance, and even if we make the questionable assumption that the construction of such probes is feasible, then silence is still the expected outcome. To see why, let's classify the possible behavioural characteristics of VN probes:

\begin{itemize}
\item{\it{Transient:}}
The probe collects information about the system, before continuing on its way to explore other systems.  \\
Result: No observable consequences

\item{\it{Persistent Non-Interference:}}
The probe remains within the system indefinitely, but in a benign manner.  \\
Result: No observable consequences based on our current technological capabilities, provided the probe is less than a few kilometres across.

\item{\it{Persistent Interference:}}
The probe remains within the system indefinitely, and exploits the available resources.
Result: No observable consequences
\end{itemize}
In other words, irrespective of the behaviour of the probe, nothing would differ from the status quo. But why is the last option deemed to be unobservable? This is yet another example of an observational selection event. In the event a probe is pathological, programmed to harvest available resources,  no primitive biological species could then  go on to evolve to the point of wresting back control of the natural resources. Therefore observers can only ever witness pristine, untouched host planets, in exactly the same way that water-based life can only witness host planets that hold liquid water. A single observation of a  water-rich host planet tells us nothing of the relative abundance of dry planets. Likewise, a single observation of an untouched host planet tells us nothing of the relative abundance of `invaded' planets.

So it is clear why we have yet to see anything, but why have we yet to \emph{hear} anything?  The above considerations don't preclude the reception of distant radio signals.  Even an optimistic application of the Drake equation \cite{drake1992anyone}, coupled with a mean longevity $L \sim 10^3$, leads to an estimate of $N_{MW} {\simlt}100$ technological civilisations in our galaxy. We should therefore fully expect to be the most populous civilisation, and perhaps by extension, also the noisiest. Detection of our nearest neighbour at ${\sim}1\,$kpc, even if they do match our electromagnetic output, is not something we should have expected to occur through serendipity. Indeed this may even fall beyond the capabilities of the Square Kilometre Array \cite{2015SKASiemion}. 

\section{Conclusions} \label{sec:conclusions}

\begin{quotation} {\itshape
Either we're gonna create simulations that are indistinguishable from reality, or civilisation will cease to exist. Those are the two options.} - Elon Musk
\end{quotation}

We have reassessed the Doomsday Argument, using Bayesian inference to account for the various possible spatial configurations of observers in the Universe. This turns out to be crucial in demonstrating why the conventional `escape route' of upweighting universes with larger populations \emph{cannot} be used to evade the Doomsday Argument. Issues relating to typicality are discussed, and we have shown that the presence of multiple reference classes is benign, provided they are selected in an unbiased manner. Our findings are found to be wholly consistent with the prediction based on frequentist reasoning. Our key conclusion is that the annual risk of global catastrophe currently exceeds $0.2\%$. In a year when Leicester City F.C. were crowned Premier League champions,  we are reminded that events of this rarity can prove challenging to anticipate, yet they should not be ignored.

Evidence presented in support of a given hypothesis will be less well received if the hypothesis in question leads to undesirable consequences. This is a well established form of cognitive bias, known as motivated reasoning \cite{kunda1990case, bastardi2011wishful}.  And what hypothesis could be more undesirable than the extinction of our species?  In light of this, we strongly advocate the use of dehumanised analogies when performing any calculation relating to the Doomsday Argument. This could help shield against cognitive biases, in a similar manner to experimental blinding techniques.

Skeptics are invited to address the following questions:
\begin{itemize}
\item{Do you consider it more likely that you have a common blood type than a rare one? \\ 
\emph{One can feign ignorance of your blood type if necessary. How could an affirmative answer  be justified unless you consider yourself to be an ordinary human?}}
\item{Why would your spatial location relative to other humans appear representative of the parent population, but not your temporal one?  \\ \emph{ Are the number of people south of your position a reasonable estimate of the total number of people north of your position? Try various reference classes: street, city, county, country, continent. Repeat for place of birth.}}
\item{Why should planets be granted a special status above other physical objects which can be used to demarcate the spatial and temporal distribution of observers  (like trains, planes, and buildings)?
\\ \emph{ The train you board tomorrow probably has more people on it than other trains running that day. Over its lifespan, it will have held roughly as many passengers after your journey, as it had before.}} 
\item{If your estimate of your personal lifespan makes use of data from other humans, aren't you already assuming typicality? \\
\emph{If so, why should an estimation of the lifespan of your species be treated  differently?}}
\end{itemize}

The connection between forecasting our civilisation's longevity and forecasting our personal longevity has been considered previously by \citet{richmond2006doomsday}.  Yet they conclude that the two cases are of a different nature because \emph{``data about lifespans is in plentiful supply"} \cite{richmond2006doomsday}. This overlooks one crucial step: in order for data about the lifespans of others to be relevant to oneself, \emph{one must assume typicality}. Thus the inferences we make of these two future lifespans - those of our personal self and that of our civilisation  - are indeed on an equal footing. We are simply more accustomed to contemplating the former. 

The consequences at stake here are not solely philosophical.  Observational selection effects can lead to a better understanding of the size and longevity of habitable planets  \cite{2015SimpsonAliens, simpson2016development}, thereby helping to optimise the search for extra-terrestrial life. Cosmological experiments can also be influenced by our vantage point \cite{2016arXiv160907120H}. But the most important implication lies closer to home. If we come to appreciate the fragility of our existence, rather than take it for granted, we are more likely to take preventative measures aimed at prolonging the future of our species.  
 
There remains one optimistic interpretation of the Doomsday Argument. The statistical constraint it imposes is on the number of future individuals, and to convert that number into a temporal duration, one must factor in the global birth rate. If technological development maintains its current pace, individual lifespans may be extended almost indefinitely. This would allow the birth rate to fall, lifting any significant bound on the longevity of our species. However even this scenario is prohibited if the appropriate reference class is ``observer moments": a segment of time experienced by each observer \cite{bostrom2013anthropic}. 

One outcome that is heavily disfavoured by the Doomsday Argument is an extended colonisation of the galaxy. This is revealed to be precisely what it sounds like: a science fiction writer's dream.  Nonetheless, more modest colonisation efforts remain feasible. For example, one's prior belief of the prospects for the successful colonisation of Mars or the Moon ought to remain intact, at least for those forecasts which spawn fewer than $10^{11}$ individuals. This is equivalent to an upper bound on the average population of around 10,000, for the remainder of the Sun's main sequence lifetime.  
  
Irrespective of the aforementioned statistical inferences, it would be na\"ive in the extreme to believe that the annual risk of global catastrophe is vanishingly small.   At a time when at least eight sovereign states are in possession of nuclear weapons (including one whose leader has executed members of his own family), a head-in-the-sand approach appears both dangerous and irresponsible. Investigations towards the mitigation of various global risks, such as those discussed by \citet{rees2004our}, ought to be pursued with urgency. We may not be able to evade the inevitable altogether, but as with our personal life expectancy, it is within our power to delay it.
\vspace{3 mm}

 \noindent{\bf Acknowledgements}\\
 The author would like to thank Antony Lewis and Alan Heavens for helpful comments, and David MacKay for bringing the topic to my attention. This work was supported by Spanish Mineco grant AYA2014-58747-P and MDM-2014-0369 of ICCUB (Unidad de Excelencia `Mara de Maeztu').

\bibliography{../../../HomeSpace/Routines/dis,../anthropic_bib}

\end{document}